\documentstyle{article}

\input{tcilatex}

\begin{document}

\vspace*{2.0cm}

\begin{center}
{\bf Noise-induced dephasing of an ac-driven Josephson junction}
\end{center}

\vspace*{1.4cm}

\begin{center}
{Giovanni Filatrella$^a$, Boris A. Malomed$^b$, and Sergio Pagano$^c$ 
\footnote{%
corresponding author. E-mail: s.pagano@cib.na.cnr.it}}

$^a$Unit\`a INFM Salerno and Facolt\`a di Scienze, Universit\`a del Sannio,
via Port'Arsa 11, 82100 Benevento, Italy

$^b$Department of Interdisciplinary Studies, Faculty of Engineering, Tel
Aviv University, Tel Aviv 69978, Israel

$^c$Istituto di Cibernetica del CNR, via Campi Flegrei 34 ed.70, I-80078
Pozzuoli (NA), Italy
\end{center}

\vspace*{2.0cm}

Josephson effects)

05.40.-a (Fluctuation phenomena, random processes, noise, and Brownian
motion)

\vspace*{2.0cm}

\begin{center}
{\bf Abstract}
\end{center}

We consider the phase-locked dynamics of a Josephson junction driven by
finite-spectral-linewidth ac current. By means of a transformation, the
effect of frequency fluctuations is reduced to an effective additive noise,
the corresponding (large) dephasing time being determined, in the
logarithmic approximation, by the Kramers' expression for the lifetime. For
sufficiently small values of the drive's amplitude, direct numerical
simulations show agreement of the dependence of the dephasing activation
energy on the ac-drive's spectral linewidth and amplitude with analytical
predictions. Solving the corresponding Fokker-Planck equation analytically,
we find a universal dependence of a critical value of the effective
phase-diffusion parameter on the drive's amplitude at a point of a sharp
transition from the phase-locked state to an unlocked one. However, for
large values of the drive amplitude, saturation and subsequent decrease of
the activation energy are revealed by simulations, which cannot be accounted
for by the perturbative analysis. The same new effect is found for a
previously studied case of ac-driven Josephson junctions with intrinsic
thermal noise. The predicted effects are relevant to applications to voltage
standards, as they determine the stability of the Josephson phase-locked
state.

\vspace*{2.0cm} \newpage

\section{Introduction}

A particle in a viscous medium, in the presence of a spatially periodic
potential, can be driven by a time-periodic force at a nonzero average
velocity $v_{0}$ determined by the resonance condition, 
\begin{equation}
2\pi m/\omega _{0}=l/v_{0},  \label{resonance}
\end{equation}
where $\omega _{0}$ is the driving frequency, an integer $m$ is the order of
the resonance, and $l$ is the period of the potential. This phenomenon was
experimentally observed in the form of phase-rotating states ({\it Shapiro
steps}) in small Josephson Junctions (JJ) driven by ac bias current \cite
{Levinsen}. Later, it was shown that the same effect can also be realized
for a fluxon (magnetic flux quantum) moving in a periodically modulated long
JJ under the action of the ac bias current \cite{we2}. The latter effect has
been recently experimentally observed, in the form corresponding to both the
fundamental resonance [$m=1$ in Eq. (\ref{resonance})] and higher-order
ones, in a long circular JJ with an effective spatial modulation induced by
uniform dc magnetic field \cite{Ustinov}.

In terms of the phase difference $\theta $ of the superconducting wave
function across the junction, an ac-driven small JJ is described by a
pendulum equation (for a detailed derivation see the book \cite{Barone})

\begin{equation}
\frac{d^{2}\theta }{dt^{2}}+\sin \theta =-\alpha \frac{d\theta }{dt}
+\epsilon \cos \left( \omega _{0}t+\psi(t) \right) +j(t)\,.\,  \label{JJ}
\end{equation}

Here, time is measured in units of the inverse Josephson plasma frequency, $%
\alpha $ is the normalized JJ conductance, $\epsilon $ is the ac-current
amplitude normalized to the Josephson critical current, $\omega _{0}$ is the
normalized driving frequency, $\psi $ is an arbitrary phase, and $j(t)$
represents intrinsic thermal noise in JJ. The equation of motion for the
fluxon in the above-mentioned long circular JJ in the magnetic field takes
exactly the same form in the ``nonrelativistic'' limit, i.e., if the
fluxon's velocity is much smaller than the Swihart velocity of the junction.
The resonant relation (\ref{resonance}), where $l=2\pi $, implies that,
neglecting the noise, the ac drive in Eq. (\ref{JJ}) may support {\em phase
rotation} of the pendulum at the average velocity $d\theta /dt\,=v_{0}$ in
the presence of the friction. Below, we assume that the spatial modulation
period is always normalized so that $l\equiv 2\pi $, i.e., $v_{0}=\omega
_{0}/m$. We note that $d\theta /dt$ is proportional to the voltage across
JJ, hence the phase rotation in the ac-driven JJ gives rise to a nonzero dc
voltage, a feature that is used in Josephson voltage standards (see, e.g.,
Ref. \cite{Pankratov} and references therein).

In real applications, the driving ac signal is always slightly
nonmonochromatic, having a finite width $\delta \omega $ in the spectral
domain; in other words, $\psi $ in Eq. (\ref{JJ}) is not a constant phase,
but rather a slowly varying function of time, representing random phase
fluctuations of the driving signal. This introduces a finite lifetime $%
\aleph $ of the phase-locked state, which is of direct relevance to
applications, affecting the stability of the Josephson voltage standards.

Intrinsic thermal fluctuations, represented by the term $j(t)$ in Eq. (\ref
{JJ}), also contribute to dephasing of the ac-driven motion. In terms of the
small ac-driven JJ, the effect of thermal fluctuations was considered in
earlier works \cite{Kautz,Ben-Jacob}, where Eq. (\ref{JJ}) with the
monochromatic drive and thermal noise was reduced to a Langevin equation for
a particle driven by a random force in a periodic potential. The
phase-locked state is then represented by the particle trapped at a minimum
of the potential, and the dephasing implies that the particle is extracted
by the random force from the trapped state. The corresponding dephasing time
was taken as the inverse Kramers' escape rate \cite{Kramers}, i.e., 
\begin{equation}
\aleph \sim \exp \left( \Delta U/T\right) ,  \label{Kramers}
\end{equation}
where $\Delta U$ is the difference between maximum and minimum values of the
effective potential, and $T$ is the temperature (see exact definitions
below).

In this work, we focus on effects of the frequency fluctuations in the ac
drive. In section 2, we demonstrate that Eq. (\ref{JJ}) with a finite
linewidth of the ac signal can be transformed, at the first order of the
perturbation theory, into an equation driven by a strictly {\em monochromatic%
} signal and an {\em additive} random force. However, in contrast to the
random force representing the intrinsic thermal noise [$j(t)$ in Eq.(\ref{JJ}%
)], the correlator of the effective additive noise generated by the
ac-drive's nonmonochromaticity does not contain the friction coefficient $%
\alpha $, as this correlator, which has a non-thermal origin, does not obey
the fluctuation-dissipation theorem. Next, using the energy-balance
technique \cite{we2}, in section 3 we reduce the monochromatically driven
equation with the random force to a Langevin equation in a periodic
potential $U(\theta )$, which makes it possible to predict the dephasing
time by means of the Kramers' expression (\ref{Kramers}).

In section 4, we specially consider the situation when the small amplitude $%
\epsilon $ of the driving signal is close to the phase-locking threshold $%
\epsilon _{{\rm thr}}$ \cite{we2}. By means of the Fokker-Planck equation
corresponding to the above-mentioned effective Langevin equation \cite
{Stratonovich,Risken}, we demonstrate that a critical value $\delta \omega $
of the driving-signal's linewidth, at which a sharp transition from the
locking to unlocking (in the form of random $2\pi $ phase slips) takes
place, may be represented as a universal function of the drive's amplitude $%
\epsilon $.

In section 5, we present results of direct numerical simulations of Eq. (\ref
{JJ}) with the nonmonochromatic drive, which are reported in the form
showing the logarithm of the dephasing time $\aleph $ as a function of $%
\epsilon $ and of the linewidth $\delta \omega $. If $\epsilon $ is above
the threshold value $\epsilon _{{\rm thr}}$, and is not too large, the
numerically found lifetime $\aleph $ is found to be quite close to that
predicted by the perturbation theory. However, at large values of $\epsilon $
the simulations reveal a new effect, which cannot be predicted by the
perturbative analysis: $\aleph (\epsilon )$ reaches a maximum value and then
decreases. As the nonmonotonic character of the dependence $\aleph (\epsilon
)$ and the existence of the maximum in it are quite important features, in
section 6 we report results of direct simulations of the mode with the
strictly monochromatic drive and intrinsic thermal (additive) noise. We
conclude that the dependence $\aleph (\epsilon )$ in this case has the same
nonmonotonic character. Although the latter model was studied earlier \cite
{Kautz,Ben-Jacob}, this feature was not reported.

It is relevant to mention that, besides small and long JJs, essentially the
same dynamical model as the one considered in this work applies to ensembles
of oscillators coupled via a mean field, which may be laser arrays or
biological oscillators \cite{strogatz94}. As is known, the global coupling
may synchronize the phase-rotation states of the oscillators, each of them
being driven by the mean field. On the other hand, various perturbations
affecting the mean field make it a slightly nonmonochromatic drive \cite
{kuramoto75,wiesenfeld92}. Thus, desynchronization of the globally coupled
rotating oscillators is another manifestation of the problem considered in
this work.

\section{Transformation of frequency fluctuations into an additive noise}

We begin our analysis from Eq. (\ref{JJ}) without intrinsic thermal noise,
i.e., with $j=0$. White-noise fluctuations of the ac-drive's frequency, $%
\omega (t)\equiv d\psi /dt+\omega _{0}$, are assumed to be subject to the
Gaussian correlations: 
\begin{equation}
\left\langle \frac{d\psi (t)}{dt}\cdot \frac{d\psi (t^{\prime })}{dt^{\prime
}}\right\rangle =2\Omega \,\delta (t-t^{\prime }),  \label{Gauss}
\end{equation}
$\Omega $ being intensity of the fluctuations. The relative linewidth of the
ac drive, $\delta \omega \equiv \left( \omega -\omega _{0}\right) /\omega
_{0}$ (calculated at $-3$ dB level), can be then estimated as $\delta \omega
\approx 2.4\,\Omega /\omega _{0}$. We note that even low-quality sources of
radiofrequency radiation, that may be used as the ac drive for JJ, have $%
\delta \omega \,\,_{\sim }^{<}\,\,10^{-3}$, while the dissipative constant $%
\alpha $ in JJ, although being small, is normally in the range $\alpha
\,\,_{\sim }^{>}\,\,10^{-2}$,$\ $therefore we hereafter assume $\delta
\omega \ll \alpha $. In other words, we may assume that a characteristic
time of the variation of the random ac-drive's phase shift $\psi (t)$ is
much larger than the relaxation time $1/\alpha $.

To convert the frequency fluctuations into an effective additive noise,
which is more convenient for the subsequent analysis, we define a new time
variable which includes a slowly varying stochastic term, 
\begin{equation}
\tau \equiv t+\chi (t){\rm ,\,with}\,\,\chi (t)\equiv \omega _{0}^{-1}\psi
(t).  \label{tau}
\end{equation}
Then, transforming the time derivatives $d/dt$ into $d/d\tau $ according to
this definition, making use of the above relations $\delta \omega \ll \alpha
\ll 1$, and keeping newly appearing small perturbations at two lowest
orders, we cast the underlying Eq. (\ref{JJ}) into a form 
\begin{equation}
\frac{d^{2}\theta }{d\tau ^{2}}+\sin \theta =-\alpha \frac{d\theta }{d\tau }
+\epsilon \cos \left( \omega _{0}\tau \right) -2\frac{d\chi }{d\tau }\cdot 
\frac{d^{2}\theta }{d\tau ^{2}}-\alpha \frac{d\chi }{d\tau }\cdot \frac{
d\theta }{d\tau }\,.\,  \label{transformed}
\end{equation}
In fact, the last term in Eq. (\ref{transformed}) is much smaller than the
previous one, as $\alpha $ is small, and, in the first approximation, one
may substitute, in the latter term, $d^{2}\theta /d\tau ^{2}\approx -\sin
\theta $. Thus, the final form of the perturbed equation, in which the
frequency fluctuations were converted into the effective additive random
force, is 
\begin{equation}
\frac{d^{2}\theta }{d\tau ^{2}}+\sin \theta =-\alpha \frac{d\theta }{d\tau }
+\epsilon \cos \left( \omega _{0}\tau \right) +2\frac{d\chi }{d\tau }\cdot
\sin \theta \,.\,  \label{final}
\end{equation}

The time transformation (\ref{tau}) affects the Gaussian correlator (\ref
{Gauss}). It is easy to find that, in terms of the renormalized time and
renormalized random phase $\chi $ (see Eq. (\ref{tau})), an exact form of
the correlator is 
\[
\left\langle \frac{d\chi (\tau )}{d\tau }\left( 1-\frac{d\chi (\tau )}{d\tau 
}\right) ^{-1}\cdot \frac{d\chi (\tau ^{\prime })}{d\tau ^{\prime }}\left( 1-%
\frac{d\chi (t^{\prime })}{d\tau ^{\prime }}\right) ^{-1}\right\rangle =%
\frac{2\Omega }{\omega _{0}^{2}}\,\frac{\delta (\tau -\tau ^{\prime })}{%
\left| 1-\frac{d\chi }{d\tau }\right| }\,.
\]
However, in view of the smallness of the frequency fluctuations, in the
lowest approximation we may adopt a simple form of the correlator, 
\begin{equation}
\left\langle \frac{d\chi (\tau )}{d\tau }\cdot \frac{d\chi (\tau ^{\prime })%
}{d\tau ^{\prime }}\right\rangle =\frac{2\Omega }{\omega _{0}^{2}}\,\delta
(\tau -\tau ^{\prime })\,.  \label{newGauss}
\end{equation}
Equations (\ref{final}) and (\ref{newGauss}) will be a basis for further
analysis, while numerical simulations will be run for the full underlying
equation (\ref{JJ}).

\section{An effective Langevin equation and estimate for the lifetime of the
phase-locked state}

In the zero-order approximation, which implies $\epsilon =\alpha =d\chi
/d\tau =0$, Eq. (\ref{final}) has a known solution, 
\begin{equation}
\theta _{0}(t)=2\,{\rm am}((\tau -\tau _{0})/k;k),  \label{am}
\end{equation}
corresponding to the phase rotation at a nonzero average frequency (phase
velocity) 
\begin{equation}
\omega _{0}=\pi /kK(k)\,.  \label{omega0}
\end{equation}
Here, {\rm am} is the Jacobi's elliptic amplitude with the modulus $k$ ($%
0<k<1$ ), $K(k)$ and $E(k)$ are complete elliptic integral of the first and
second kinds, and $\tau _{0}$ is an arbitrary constant. In this
approximation, Eq. (\ref{JJ}) conserves the energy, the value of which for
the law of motion (\ref{am}) is determined by the value of $k$, 
\begin{equation}
{\cal E}\equiv \frac{1}{2}\left( \frac{d\theta _{0}}{d\tau }\right)
^{2}-\cos \theta _{0}=\frac{2}{k^{2}}-1\,.  \label{energy}
\end{equation}

A possibility to support persistent phase rotation at a velocity $\omega _{0}
$ in the presence of the friction by the monochromatic ac drive is predicted
by the energy-balance equation \cite{we2}. To this end, we calculate the net
rate of the change of energy due to the action of the friction and drive,
averaged over the rotation period $2\pi /\omega _{0}$, under the resonance
condition given by Eq. (\ref{resonance}): 
\begin{equation}
\overline{\frac{d{\cal E}}{d\tau }}=-\alpha \,\overline{\left( \frac{d\theta
_{0}}{d\tau }\right) ^{2}}+\epsilon \omega _{0}\,\omega _{1}\cos (\omega
_{0}\tau _{0}),  \label{oldbalance}
\end{equation}
where the overbar stands for the time average, $\tau _{0}$ being the same
constants as in Eq. (\ref{am}), and $\omega _{1}$ is the amplitude of the
resonant harmonic in the Fourier decomposition of the time-dependent
velocity (instantaneous frequency) $d\theta _{0}/d\tau $, taken as per the
unperturbed law of motion (\ref{am}). An elementary calculation yields 
\begin{equation}
\overline{\left( \frac{d\theta _{0}}{d\tau }\right) ^{2}}=\frac{4E(k)}{%
k^{2}K(k)}\,,\,\,\omega _{1}=\frac{2Q}{1+Q^{2}},\,\,Q\equiv \exp \left[ -\pi
K\left( \sqrt{1-k^{2}}\right) /K\right]   \label{Jacobi}
\end{equation}
($Q$ is called the Jacobi parameter). Then, the phase-locked ac-driven
regime, corresponding to $\stackrel{\_\_\_\_\_\_\_}{d{\cal E}/d\tau }$ $=0$,
is possible at two constant values (one stable and one unstable) of the
phase difference between the ac drive and the rotating pendulum, 
\begin{equation}
\omega _{0}\tau _{0}=\pm \cos ^{-1}\left[ \frac{\alpha }{\epsilon \omega
_{0}\omega _{1}}\,\overline{\left( \frac{d\theta _{0}}{d\tau }\right) ^{2}}%
\right] \,  \label{phaseshift}
\end{equation}
\cite{we2}, provided that the amplitude $\epsilon $ exceeds the threshold
value 
\begin{equation}
\epsilon _{{\rm thr}}={\rm \ }\frac{\alpha }{\omega _{0}\omega _{1}}\,%
\overline{\left( \frac{d\theta _{0}}{d\tau }\right) ^{2}}\,.
\label{threshold}
\end{equation}

In the presence of the additive random force in Eq. (\ref{final}), a
perturbed equation of motion for the rotating pendulum can again be obtained
from the averaged energy-balance equation, which has the same form of Eq. (%
\ref{oldbalance}) with a difference that now $\tau _{0}$ is a slowly varying
function of the time $\tau $ [roughly speaking, varying as slowly as the
random phase $\psi (t)$ in the underlying equation (\ref{JJ})]. Notice that
the time dependence of $\tau _{0}$ determines a change $\delta \omega
_{0}=\omega _{0}\left( d\tau _{0}/d\tau \right) $ of the average
phase-rotation velocity, and the latter may be related to the change of the
energy (\ref{energy}), through its kinetic part, as $\delta {\cal E}={\cal E}%
^{\prime }\delta \omega _{0}$, where ${\cal E}^{\prime }$ stands for $d{\cal %
E}/d\omega _{0}$, calculated for the unperturbed law of motion given by Eq. (%
\ref{am}), ${\cal E}^{\prime }=4K^{2}/[\pi k(K+kdK/dk)]$). Thus, $\stackrel{%
\_\_\_\_\_\_\_}{d{\cal E}/d\tau }={\cal E}^{\prime }\omega _{0}\left(
d^{2}\tau _{0}/d\tau ^{2}\right) $, and the balance equation (\ref
{oldbalance}) takes the form 
\begin{equation}
\omega _{0}{\cal E}^{\prime }\frac{d^{2}\tau _{0}}{d\tau ^{2}}\,=-\alpha \,%
\overline{\left( \frac{d\theta _{0}}{d\tau }\right) ^{2}}(1+2\frac{d\tau _{0}%
}{d\tau })+\epsilon \omega _{0}\omega _{1}\cos \left( \omega _{0}\tau
_{0}\right) +2\omega _{0}\frac{d\chi }{d\tau }\cdot \sin \left( \theta (\tau
)\right) ,  \label{newbalance}
\end{equation}
the second term in $(1+2d\tau _{0}/d\tau )$ being a contribution to the
energy dissipation rate due to the small change of the average velocity,
while a similar correction to the last term in Eq. (\ref{newbalance}) is
negligible. Eq. (\ref{newbalance}) can be transformed into a more convenient
form by defining $\omega _{0}\tau _{0}(\tau )\equiv \zeta (\tau )\,$: 
\begin{equation}
\omega _{0}{\cal E}^{\prime }\frac{d^{2}\zeta }{d\tau ^{2}}+\,2\alpha 
\overline{\left( \frac{d\theta _{0}}{d\tau }\right) ^{2}}\,\frac{d\zeta }{%
d\tau }=[-\omega _{0}\alpha \overline{\left( \frac{d\theta _{0}}{d\tau }%
\right) ^{2}}+\epsilon \omega _{0}^{2}\omega _{1}\cos \zeta ]+2\omega
_{0}^{2}\sin \left( \theta (\tau )\right) \frac{d\chi }{d\tau }.
\label{Langevin1}
\end{equation}

Thus, we have arrived at an effective Langevin equation \cite{Risken} for a
particle driven in a viscous medium by the sum of a regular force,
represented by the terms in the square brackets, and a stochastic force,
represented by the last term in the equation. The subsequent step is,
following a well-known procedure \cite{Stratonovich,Risken}, to introduce
the Fokker-Planck (FP) equation corresponding to this Langevin equation,
taking into regard the correlator (\ref{newGauss}). An essential feature of
the thus derived FP equation is the presence of the extra multiplier $\sin
^{2}\left[ \theta (\tau )\right] $ in front of the diffusion
(second-derivative) term in it, due to the multiplier $\sin \left[ \theta
(\tau )\right] $ in the stochastic-force term in Eq. (\ref{Langevin1}). The
coefficient $\sin ^{2}\left[ \theta (\tau )\right] $ may be averaged in time
as per the unperturbed law of motion (\ref{am}). It is easy to calculate the
average value: 
\begin{equation}
\overline{\sin ^{2}\theta _{0}(\tau )}=\frac{4}{3k^{4}}\left[ \left(
2-k^{2}\right) \frac{E(k)}{K(k)}-2\left( 1-k^{2}\right) \right] 
\label{average}
\end{equation}
[note that, in the limit $k\rightarrow 0$, the expression (\ref{average})
approaches an obvious value $1/2$].

The FP equation takes essentially the same form as it would take in the
known problem \cite{Kautz,Ben-Jacob} of the dephasing of the ac-driven JJ
phase rotation under the action of the additive thermal noise represented by
the term $j(t)$ in Eq. (\ref{JJ}) [the most important difference of the
effective Langevin equation (\ref{Langevin1}) from its counterpart in the
thermal-noise problem is the presence of the multiplier $\sin \left[ \theta
(\tau )\right] $ in the last term of the equation]. With the FP equation
taking the usual form, one can directly use the Kramers' expression (\ref
{Kramers}) to determine the lifetime of the ac-driven state, as it was done
for the case of the thermal noise in Refs. \cite{Kautz,Ben-Jacob}.\ In
particular, an effective potential corresponding to the potential force,
i.e., the expression in square brackets in Eq. (\ref{Langevin1}), is 
\[
U(\zeta )=-\epsilon \omega _{0}^{2}\omega _{1}\sin \zeta +\omega _{0}\alpha 
\overline{\left( \frac{d\theta _{0}}{d\tau }\right) ^{2}}\zeta ,
\]
hence the potential-barrier height $\Delta U$, which should be substituted
into Eq. (\ref{Kramers}), can be easily found as a difference of values of
the potential taken between two points where the above-mentioned potential
force vanishes. As a result, we obtain 
\begin{equation}
\Delta U=2\epsilon \omega _{0}^{2}\omega _{1}\left[ \sqrt{1-\left( \epsilon
_{{\rm thr}}/\epsilon \right) ^{2}}-\left( \epsilon _{{\rm thr}}/\epsilon
\right) \cos ^{-1}\left( \epsilon _{{\rm thr}}/\epsilon \right) \right] \,,
\label{DeltaU}
\end{equation}
where the definition (\ref{threshold}) for the threshold value of the
amplitude has been used to simplify the expression.

However, an additional difference of the present case from the thermal-noise
problem, that must be taken into regard before applying the expression (\ref
{Kramers}), is that the frequency fluctuation intensity appearing in the
correlator (\ref{newGauss}) does {\em not} obey the fluctuation-dissipation
theorem, and therefore it does not include the intrinsic dissipative
constant $\alpha $ of the pendulum (JJ). By properly defining the effective
temperature $T_{{\rm eff}}=4\omega _{0}^{2}\,\overline{\sin ^{2}\theta
_{0}(\tau )}\,\Omega /\alpha _{{\rm eff}}$, where $\alpha _{{\rm eff}}\equiv
2\alpha \overline{\left( d\theta _{0}/d\tau \right) ^{2}}$ is the effective
friction constant from Eq.~(\ref{Langevin1}), and using the
potential-barrier height $\Delta U$ (\ref{DeltaU}) and the expression (\ref
{average}), we can rewrite the Kramers' expression (\ref{Kramers}) as: 
\begin{equation}
\aleph \sim \exp \left( \frac{6\epsilon \alpha k^{2}EQ}{\left[ \left(
2-k^{2}\right) E-2\left( 1-k^{2}\right) K\right] \left( 1+Q^{2}\right)
\Omega }\left[ \sqrt{1-\left( \epsilon _{{\rm thr}}/\epsilon \right) ^{2}}
-\left( \epsilon _{{\rm thr}}/\epsilon \right) \cos ^{-1}\left( \epsilon _{%
{\rm thr}}/\epsilon \right) \right] \right) \,.  \label{main}
\end{equation}
This is the main prediction of the analytical consideration, which will be
compared to results of direct simulations of Eq. (\ref{JJ}) in section 5.

\section{Dephasing the phase-locked state near the locking threshold}

In this section we investigate the system described by the Langevin equation
(\ref{Langevin1}), by explicitly solving the associated FP equation.
However, we can first simplify Eq. (\ref{Langevin1}), recalling the
fundamental physical condition according to which the frequency fluctuations
are much smaller than $\alpha $, or, in other words, the random force varies
on a time scale $\gg 1/\alpha $ . Consequently, the acceleration term on the
left-hand side of Eq. (\ref{Langevin1}) may be neglected as compared to the
velocity term, which yields a simplified Langevin equation

\begin{equation}
\dot{\zeta }\,=-F_{0}+F_{1}\cos \zeta +f(t),  \label{Langevin2}
\end{equation}
where $F_{0}\equiv \omega _{0}/2$, $F_{1}\equiv \epsilon \omega
_{0}^{2}\omega _{1}[2\alpha \overline{\left( d\theta _{0}/dt\right) ^{2}}
]^{-1}$, and 
\begin{equation}
f(t)\equiv \omega _{0}^{2}\left[ \alpha \,\overline{\left( \frac{d\theta
_{0} }{d\tau }\right) ^{2}}\right] ^{-1}\frac{d\chi }{d\tau }\cdot \sin %
\left[ \theta (\tau )\right] .  \label{f}
\end{equation}

The FP equation (in this case, it is, in fact, the Smoluchowski equation 
\cite{Risken,Stratonovich}) for a probability distribution function $P(\zeta
,t)$, corresponding to Eq. (\ref{Langevin2}) with the Gaussian correlator (%
\ref{newGauss}), is 
\begin{equation}
P_{t}=F_{1}\sin \zeta \cdot P+\left( F_{0}-F_{1}\cos \zeta \right) P_{\zeta
}+\Xi P_{\zeta \zeta }\,,  \label{FP}
\end{equation}
where the subscripts stand for the partial derivatives, and 
\begin{equation}
\Xi \equiv \,\overline{\sin ^{2}\theta _{0}(\tau )}\,\left[ \alpha \,%
\overline{\left( \frac{d\theta _{0}}{d\tau }\right) ^{2}}\right]
^{-2}\,\omega _{0}^{2}\,\Omega \,,  \label{Xi}
\end{equation}
where $\Omega $ is the same as in Eqs. (\ref{Gauss}) and (\ref{newGauss}),
and the average value $\overline{\sin ^{2}\theta _{0}(\tau )}$ is given by
Eq. (\ref{average}).

Information about the distribution of the phase $\zeta $ can be obtained
from the stationary version of Eq. (\ref{FP}), 
\begin{equation}
D\frac{d^{2}P}{d\zeta ^{2}}=-(1-b\cos \zeta )\frac{dP}{d\zeta }-b\sin \zeta
\cdot P\,,  \label{statFP}
\end{equation}
where the final set of notation is $b\equiv F_{1}/F_{0}$ and $D\equiv \Xi
/F_{0}$. These two parameters are interpreted , respectively, as the ratio
of the drive's amplitude to the friction coefficient, and an FP diffusion
coefficient, which is proportional to an effective drive's linewidth.

In the absence of the diffusion ($D=0$), a solution to Eq. (\ref{statFP}) is 
\begin{equation}
P(\zeta )=\left( 2\pi \right) ^{-1}\sqrt{1-b^{2}}/(1-b\cos \zeta ),
\label{simplesolution}
\end{equation}
where the normalization $\int_{-\pi }^{+\pi }P(\zeta )d\zeta =1$ is imposed.
The solution (\ref{simplesolution}) is regular at $b<1$, while the
singularity at $b=1$ exactly corresponds to the drive's amplitude attaining
the threshold value (\ref{threshold}), i.e., to the onset of the
phase-locking regime.

Collecting results produced by the numerical solution of Eq. (\ref{statFP}),
we have concluded that it is possible to define a {\em critical value} $D_{%
{\rm cr}}$ of $D$, at which a sharp transition from the phase-locked state
at $D<D_{{\rm cr}}$ to an unlocked one at $D>D_{{\rm cr}}$ takes place. The
transition is still better illustrated by consideration of the probability
flux, $J\equiv $ $-\,\left[ DP^{\prime }+(1-b\cos \zeta )P\right] $, in
terms of which the time-dependent FP (Smoluchowski) equation (\ref{FP}) is
written as $F_{0}^{-1}P_{t}+$ $\partial J/\partial \zeta =0$. At the points
of a minimum of the stationary distribution function, $|J|$ gives the
phase-slippage rate, i.e., a rate of the transition from a vicinity of a
phase-locked point to a point differing by a phase shift $\pm 2\pi $. In
Fig. 1, we display the phase-slippage rate vs. $D$ at different constant
values of $b$, as found from the numerical solution of Eq. (\ref{statFP}).
The existence of critical values $D_{{\rm cr}}$, such that virtually no
phase slippage takes place at $D<D_{{\rm cr}}$, is evident. Note that for $%
b<1$, when the locking is impossible even to the monochromatic drive in the
absence of the noise ($D=0$), $D_{{\rm cr}}$ does not exist. For $b=1$,
i.e., exactly at the locking threshold (\ref{threshold} ), $D_{{\rm cr}}=0$,
which cannot be seen on the logarithmic scale used in Fig. 1. It is
necessary to mention that a picture which may be interpreted as showing the
flux $J$ as a function of $b$ at different fixed values of $D$ is available
in the book \cite{Risken}; nevertheless, we find it relevant to present Fig.
1 here, as we need to display $J(D)$ at different fixed values of $b$
(otherwise the existence of the critical values $D_{{\rm cr}}$ is not
obvious).

The sharp unlocking transition can also be seen in terms of the ratio of the
aforementioned minimum value of $P(\zeta )$ to its maximum value, which is
attained fairly close to the unperturbed (i.e., pertaining to the
monochromatic drive) locking point. These data (not displayed here) show
that the ratio is virtually equal to zero at $D<D_{{\rm cr}}$, and abruptly
begins to increase exactly at $D=D_{{\rm cr}}$. In order to quantify $D_{%
{\rm cr}}$ we define it as the value of $D$ for which $J=10^{-3}$.

Figure 2 shows the most important characteristic of the unlocking
transition, viz., $D_{{\rm cr}}$ vs. the effective drive's amplitude $b$.
The dependence is nearly linear, except for the region $0<b-1\ll 1$, i.e.,
just above the locking threshold (\ref{threshold}). It is difficult to show
the dependence in this region directly, therefore, in Fig. 2 we instead
display $r(b)$ for $b-1\ll 1$, where $r$ is defined so that $D_{{\rm cr}}$
is approximated by an expression $C_{1}(b-1)^{r}$ with a suitable constant $%
C_{1}$. 

An asymptotic value of $r$ for $b-1\rightarrow 0$ can be found analytically,
as one can use the exact solution (\ref{simplesolution}) to describe an
approximate form of the distribution function, except in a sensitive region
of small $\zeta $: $P(\zeta )\sim 1/\sin ^{2}(\zeta /2)$. At small $\zeta $,
one can expand Eq. (\ref{statFP}), taking into regard that $b-1$ and $D$ are
now small too. This yields an equation, 
\begin{equation}
\frac{d^{2}P}{d\eta ^{2}}=-\left( \frac{1}{2}\eta ^{2}-\frac{b-1}{D^{2/3}}%
\right) \frac{dP}{d\eta }-\eta P,  \label{expanded}
\end{equation}
where $\eta \equiv D^{-1/3}\zeta $. Thus, the solution in the sensitive
region (which is $\eta \sim 1$, or $\zeta \sim \sqrt{b-1}\sim D^{1/3}$)
depends on the single parameter $(1-b)/D^{2/3}$. Although a solution to Eq. (%
\ref{expanded}) can be matched to the aforementioned approximation $P(\zeta
)\sim 1/\sin ^{2}(\zeta /2)$, valid at larger $\zeta $, only numerically, it
is obvious that, as $b-1\rightarrow 0$, the dependence $D_{{\rm cr}}(b)$
must be $D_{{\rm cr}}=C_{1}(b-1)^{3/2}$, with $C_{1}\approx 0.17$ found from
numerical data. The value $r=3/2$, obtained for $b-1\rightarrow 0$, is in
good agreement with the numerical results displayed in Fig. 2.

The linearity of the dependence $D_{{\rm cr}}(b)$ at large $b$ can be
explained in a very simple way: neglecting in this case $1$ in the
expression $(1-b\cos \zeta )$ in Eq.~(\ref{statFP}), we immediately conclude
that the asymptotic solution depends on the single parameter $b/D$, hence
the dependence must take the form $D_{{\rm cr}}=C_{2}b$, with a constant $%
C_{2}\approx 0.14$ found numerically. The latter result can also be
interpreted in another way: the minimum (threshold) value of the ac-drive's
amplitude necessary to support the rotation of the pendulum grows nearly
linearly with the linewidth $\delta \omega $, so that it may be approximated
by $\epsilon _{{\rm thr}}\left( \delta \omega \right) \approx \epsilon _{%
{\rm thr}}^{(0)}(1+{\rm const\cdot }\delta \omega $), where $\epsilon _{{\rm %
thr}}^{(0)}$ is given by Eq.~(\ref{threshold}) and the constant is roughly $%
1/C_{2}$.

These analytical results, which comply well with the numerical findings,
justify the introduction of the very concept of the critical value $D_{{\rm %
cr}}$ of the phase-diffusion constant in the Smoluchowski equation, which
was originally defined above in a phenomenological way, just by looking at
Fig. 1.

\section{Numerical results}

To check the limits of validity of the analytical results obtained above, we
have performed numerical simulations of the full stochastic equation (\ref
{JJ}), using a simple Euler scheme (the use of this scheme is considered in
the book \cite{Risken}). As usual, we halved the time step until the results
would converge to a stable value within few-per-cent accuracy. In this
section, we will first focus on the case of the parametric noise, so we now
set $j(t)=0$, keeping the random term $\psi (t)$ in Eq. (\ref{JJ}). The
basic phenomenon sought in the previous analysis was the escape from the
state synchronized with the external drive as per Eq. (\ref{resonance}).
However, since the JJ is driven by the ac-term alone, once the system is no
longer phase-locked, it cannot sustain progressive motion and will therefore
quickly decay to the zero-voltage state. So the prediction of Eq.(\ref{main}
) actually refers to the lifetime of the phase-locked state; an abrupt
transition from this state to the zero-voltage one (after about $300$ time
unit) is evident in Fig. 3, which displays the time dependence of the phase
velocity, found in a typical run of the simulations of the stochastic
equation (\ref{JJ}).

To estimate the lifetime of the phase-locked state at a given
``temperature'' (spectral linewidth of the drive), we have run the
simulations for many different realizations of the random phase $\psi (t)$,
and averaged the results for the lifetime. The number of the realizations
was determined by the condition that the computed average has to converge to
an established value. In Fig. 4, the logarithm of the thus computed average
lifetime is plotted versus the inverse linewidth, so that the potential
barrier (\ref{DeltaU}) could be estimated from the slope of the linear part
of this dependence.

This method closely follows that of Ref. \cite{Kautz}; the main difference
is, as already mentioned, that we are not looking for mere phase-slippage,
but for a jump to the state with zero average velocity. As it is clearly
seen in Fig. 3, this occurs at a somewhat later time than the phase slips
commence, although the difference is, typically, small (for instance, it is
less than $20$ time units in the example shown in Fig. 3).

The numerically computed effective energy barrier (represented by lines with
symbols) and the one predicted by Eq. (\ref{DeltaU}) (the lines without
symbols) are shown in Fig. 5(a). Taking into regard that no fitting
parameter was employed, the agreement is very good near the threshold value (%
\ref{threshold}), which is $\epsilon _{{\rm thr}}\approx 0.16$ for the
values of the parameters corresponding to Fig. 5(a). However, a drastic
deviation from the analytical prediction is evident at larger values of the
ac-drive's amplitude: while the analytical formula (\ref{main}) predicts an
almost linear increase of the effective barrier height with the ac drive
amplitude, the numerical results show a maximum followed by a substantial
decrease of barrier's height. 

It is relevant to mention that the nearly linear increase of the barrier
height with $\epsilon $ was predicted by the power-balance approach, which
was employed above for the analytical consideration (see also Ref. \cite{we2}%
), a different method, based on  an harmonic expansion, would result in a
nonmonotonic dependence of the barrier height on $\epsilon $ \cite
{Ben-Jacob,Kautz}. In the case of the additive noise, the latter method
produced the energy barrier demonstrating a Bessel-functional behavior,
typical of the rf-induced current steps in JJs \cite{Ben-Jacob,Kautz}. In
our case, however, such a dependence cannot be analytically justified.
Moreover, even if the results of Ref. \cite{Kautz} are formally applied to
our case, yielding 
\begin{equation}
\Delta U\approx J_{1}\left( \epsilon /\omega _{0}^{2}\right) \,,
\label{bessel}
\end{equation}
where $J_{1}$ is the Bessel function, the maximum of $\Delta U$ would occur
at a much larger value of $b$ [$b\approx 45$, instead of $4$ in Fig. 5(a)
for $\alpha =0.01$]. Thus, the phenomenon reported here is an essentially
new one, and still remains to be explained.

It should, moreover, be noticed that the Bessel-function approximation
similar to Eq. (\ref{bessel}) gives, according to Ref. \cite{Kautz}, a good
estimate for the energy barrier only in the limit $\omega _{0}^{-2}<<1$,
deviations occurring already for $\omega _{0}^{-2}\simeq 0.05$. Therefore,
it is not surprising that, for the parameters considered here ($\omega
_{0}^{-2}=0.25$), the agreement with the Bessel-function behavior is very
poor.

To check that the newly found dependence of the effective barrier height vs.
the normalized drive's amplitude is not due to some particular feature of
the parametric noise, we have also performed extensive simulations of the
same stochastic equation (\ref{JJ}), but with the additive noise and
strictly monochromatic driving signal, for the same values of parameters as
those used above in the case of the frequency fluctuations. A typical
example is shown in Fig. 5(b), together with the theoretical estimate of the
energy barrier according to Ref.~\cite{Ben-Jacob,Kautz}:

\begin{equation}
\Delta U = 2 J \left( b \frac{ \epsilon_{{\rm thr}} }{\omega_0^2} \right) %
\left[ \sqrt{1- b^{-2} } - b^{-1} \cos ^{-1} \left( b^{-1} \right) \right]
\label{theoradd}
\end{equation}

In this case too, a strong deviation of $\Delta U(b)$ from the linear
dependence occurs at relatively low values of the drive amplitude ($b\simeq 7
$), although they are higher than those in the frequency-noise case [which
are $b\simeq 4$, see Eq. 5(a)].

\section{Conclusions}

Results reported in this work are of relevance for applications to systems
in which Josephson junctions are phase-locked to an external ac source, such
as voltage standards. By substituting reasonable experimental values into
Eq. (\ref{main}), we can estimate that an ac source with the relative
linewidth better than $10^{-4}$ is needed if the lifetime of the order of $1$
second is required for the measurement system.

We also note that our approach could be applied to another problem: a
pendulum driven by an ac signal whose frequency is subject to a systematic
(rather than random) change, i.e., a zero-linewidth but variable-frequency
drive. A system of the latter type was considered in Ref.~\cite{friedland98}
for a soliton in a perturbed nonlinear Schr\"{o}dinger equation.

This work was performed in the framework of the bilateral-cooperation
agreement between Consiglio Nazionale di Ricerca (Italy) and the Israeli
Ministry of Science and Technology, under the project Nonlinear Dynamics of
Josephson Networks. The authors appreciate useful discussions with G.
Costabile and access to computer facilities at the University of l'Aquila
(Italy).

\vspace*{2.0cm} 

\newpage \vspace*{2.0cm}

\section*{Figure Captions}

Fig. 1. The phase-slippage rate $J$ vs. the diffusion parameter $D$ at
different fixed values of the normalized drive's amplitude $b$, which are
indicated near each curve. \newline

\noindent Fig. 2. The normalized noise threshold $D_{cr}$ vs. the normalized
drive's amplitude $b$ (solid line). The dashed line shows the dependence of
the exponent $r$ defined in the text. \newline

\noindent Fig. 3. A typical example of the evolution of the phase velocity $%
\dot{\theta}(t)$ obtained from the numerical integration of the stochastic
equation (\ref{JJ}). A loss of phase-locking occurs at $t\simeq 300$. The
parameters are $\alpha =0.01$, $\epsilon =0.2$, $\Omega =0.01$, $\omega
_{0}=2$. \newline

\noindent Fig. 4. Lifetime of the phase-locked state, plotted on a
logarithmic scale versus the inverse linewidth. The parameters are $\alpha
=0.01$, $\omega _{0}=2$. The estimated slope is $0.00031$ for $\epsilon =0.2$%
, and $0.0012$ for $\epsilon =0.5$. \newline

\noindent Fig. 5. Dependence of the normalized energy barrier $\Delta U$ on
the normalized drive's amplitude $b$, for the case of frequency fluctuations
(a) and additive noise (b). Symbols represent the energy barrier estimated
numerically on the basis of Eq. (\ref{Kramers}) for $\omega _{0}=2$ and $%
\alpha =0.1$ (squares) and $\alpha =0.01$ (triangles), connecting lines
being a guide for the eye. The curves without symbols represent analytical
predictions, viz., Eq.~(\ref{main}) for the frequency-fluctuation case [in
the panel (a), the continuous and dashed curves pertain, respectively, to $%
\alpha =0.1$ and $\alpha =0.01$], and Eq.~(\ref{theoradd}) for the
additive-noise case.

\end{document}